\documentclass[%
 reprint,
superscriptaddress,
 amsmath,amssymb,
 aps,
]{revtex4-2}

	\usepackage{bm}
	\usepackage{braket}
	\usepackage{dsfont}

 	\usepackage{graphicx}
	\usepackage[caption=false]{subfig}
	\usepackage[export]{adjustbox}

	\newcommand{\vect}[1]{\boldsymbol{#1}}		
	\newcommand{\op}[1]{\hat{\boldsymbol{#1}}}	
	

\begin{document}

\title{Valley-Exchange Coupling Probed by Angle-Resolved Photoluminescence}

\author{Joshua J. P. Thompson}
\email{thompson@chalmers.se}
\affiliation{ Department of Physics, Chalmers University of Technology, 412 96 Gothenburg, Sweden}
\author{Samuel Brem}
\affiliation{Department of Physics, Philipps-Universit\"{a}t Marburg, Renthof 7, 35032 Marburg}
\author{Hanlin Fang}
\affiliation{Department of Microtechnology and Nanoscience (MC2), Chalmers University of Technology, 412 96 Gothenburg, Sweden}
\author{Carlos Ant\'{o}n-Solanas}
\affiliation{Institute of Physics, University of Oldenburg, 26129 Oldenburg, Germany}
\author{Bo Han}
\affiliation{Institute of Physics, University of Oldenburg, 26129 Oldenburg, Germany}
\author{Hangyong Shan}
\affiliation{Institute of Physics, University of Oldenburg, 26129 Oldenburg, Germany}
\author{Saroj P. Dash}
\affiliation{Department of Microtechnology and Nanoscience (MC2), Chalmers University of Technology, 412 96 Gothenburg, Sweden}
\author{Witlef Wieczorek}
\affiliation{Department of Microtechnology and Nanoscience (MC2), Chalmers University of Technology, 412 96 Gothenburg, Sweden}
\author{Christian Schneider}
\affiliation{Institute of Physics, University of Oldenburg, 26129 Oldenburg, Germany}
\author{Ermin Malic}
\affiliation{ Department of Physics, Chalmers University of Technology, 412 96 Gothenburg, Sweden}
\affiliation{Department of Physics, Philipps-Universit\"{a}t Marburg, Renthof 7, 35032 Marburg}

\date{\today}

\begin{abstract}
The optical properties of monolayer transition metal dichalcogenides are dominated by tightly-bound excitons. They form at distinct valleys in reciprocal space, and can interact via the valley-exchange coupling, modifying  their dispersion considerably. Here, we predict that angle-resolved photoluminescence can be used to probe the changes of the excitonic dispersion. The exchange-coupling leads to a unique angle dependence of the emission intensity for both circularly and linearly-polarised light. We show that these emission characteristics can be strongly tuned by an external magnetic field due to the valley-specific Zeeman-shift. We propose that angle-dependent photoluminescence measurements involving both circular and linear optical polarisation as well as magnetic fields should act as strong verification of the role of valley-exchange coupling on excitonic dispersion and its signatures in optical spectra. 
\end{abstract}

\maketitle

 Atomically thin transition metal dichalcogenides (TMDs) possess unique optical and electronic properties \cite{manzeli20172d, mak2010atomically, splendiani2010emerging}. One of the most fundamental of these is the prevalence of stable 2D excitons with large binding energies \cite{wang2018colloquium,mueller2018exciton, berghauser2018mapping} in mono- to few-layer samples. The strong optical response of  excitons provides an ideal platform to explore spintronics\cite{xiao2012coupled,nagaosa2010anomalous}, valleytronics\cite{schaibley2016valleytronics,lee2016electrical, thompson2019valley}, single-photon emitters \cite{tonndorf2015single, palacios2016atomically,thompson2020criteria}, molecule-detection \cite{feierabend2017proposal}, moir\'{e} physics \cite{tang2020simulation, brem2020tunable} and topological properties \cite{wu2017topological}.   
 
 The optical response, non-equililbrium dynamics and diffusion behaviour of TMDs is strongly determined by the excitonic band structure \cite{selig2016excitonic, brem2018exciton, perea2019exciton, gupta2019fundamental}. Hence, understanding and probing the exciton dispersion becomes of paramount importance.  For a typical photoluminescence (PL) experiment, high numerical aperture detectors are used to increase the collection/detection efficiency \cite{ splendiani2010emerging, mak2010atomically}, which leads to the simultaneous detection of photons emitted at a range of angles. Therefore, as a consequence of the exciton-light interaction, many points on the excitonic dispersion are  probed simultaneously, preventing the excitonic dispersion from being resolved. Instead, angle-resolved photoluminescence \cite{schneider2019shedding} (ar-PL) could be used to extract the energy-momentum relation of excitons in a monolayer TMD within the light cone, requiring non-orthogonal photo-detection \cite{wang2017plane}. While experimentally more challenging than standard PL, refined detection setups have been implemented improving the signal-to-noise ratio \cite{schneider2019shedding,hanisch2020refined}. Other methods to extract the excitonic dispersion such as ARPES, while successful in capturing the signatures of momentum-dark excitons \cite{madeo2020directly, christiansen2019theory, wallauer2021momentum, dong2021direct}, typically have a much lower momentum and energy resolution than angle-resolved PL measurements \cite{schneider2019shedding, schneider2020optical} and too low to resolve the valley-exchange coupling we discuss here. 
 
 In a typical monolayer TMD,  excitons form at the two distinct valleys $K$ and $K'$, each carrying a well-known circular optical polarisation and spin dependence \cite{xiao2012coupled, schaibley2016valleytronics, berghauser2018inverted}. Charge carriers in these valleys can interact both indirectly via phonons \cite{christiansen2017phonon,brem2020phonon} or directly through the Coulomb interaction \cite{schmidt2016ultrafast, yu2014dirac, qiu2015nonanalyticity, deilmann2019finite, guo2019exchange, selig2020suppression, erkensten2021exciton}. In particular, it has been predicted that the valley-exchange interaction leads to a stark reconstruction of the exciton band structure giving rise to a splitting into a parabolic-like and linear-like dispersion \cite{yu2014dirac, qiu2015nonanalyticity}. In addition, the valley-exchange coupling in TMD monolayers has been proposed as a primary source of valley depolarisation in these materials \cite{yu2014dirac, lagarde2014carrier, hao2016direct}, crucial for valleytronics.
 
 Here, we predict the angle-resolved PL of an MoSe$_2$ monolayer based on a fully microscopic many-particle theory. We find strong signatures in the  PL spectrum, which arise as a result of the valley-exchange interaction. We calculate on microscopic footing the linewidth of exciton resonances in presence of the exchange-coupling and demonstrate that the band splitting can be resolved for temperatures up to 100 K. Furthermore, we show that an external magnetic field allows the valley polarisation of each band to be tuned continuously as a function of emission angle, and that this is directly reflected in angle-resolved PL spectra.

  \begin{figure} [t!]
    \centering
    \includegraphics[width = 0.95\linewidth]{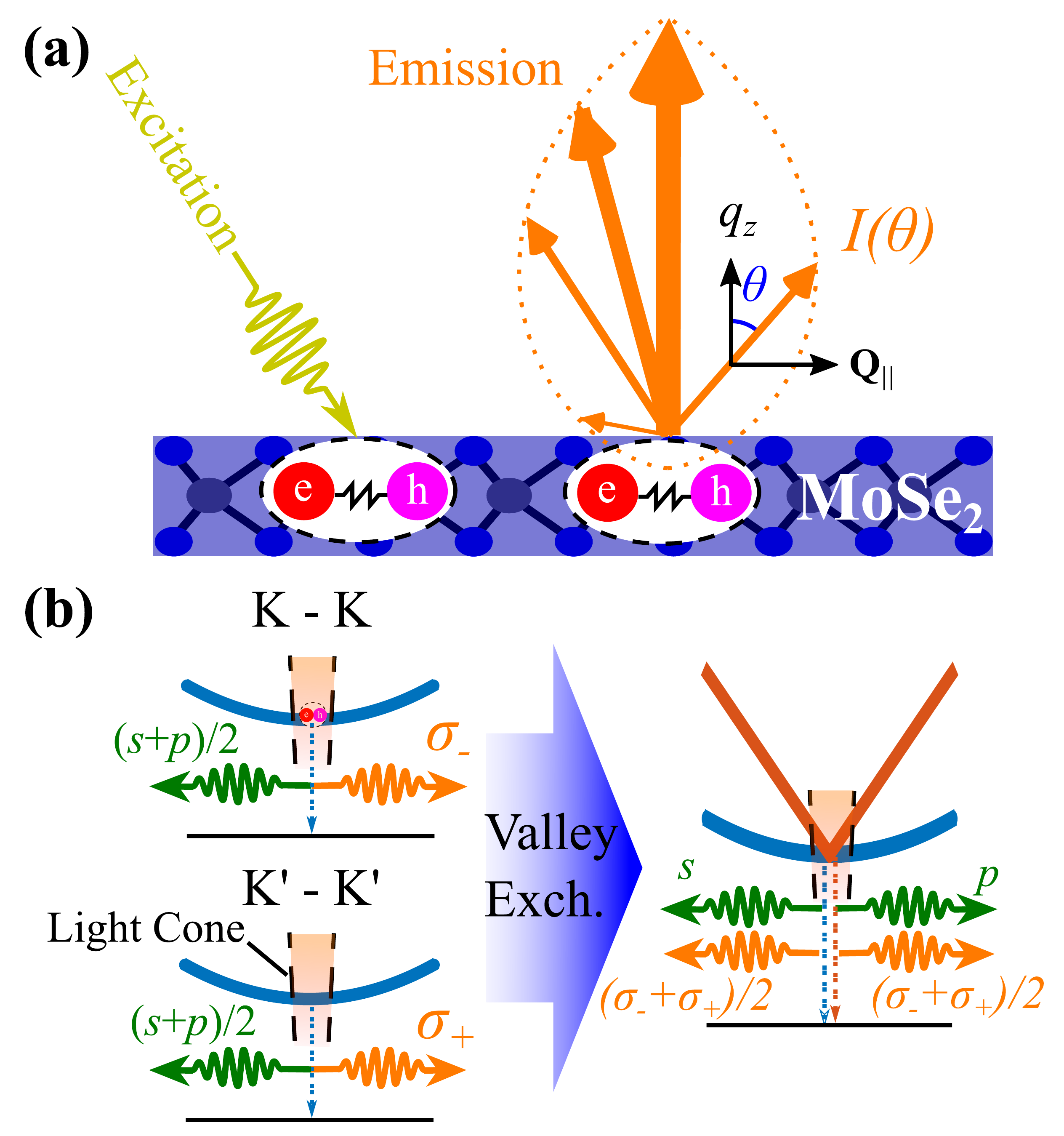}
    \caption{(a) Schematic of typical angle-resolved photoluminescence. Optical excitation leads to the formation of excitons within the MoSe$_2$ layer. After relaxation, excitons emit photons (orange arrows) with direction determined by the excitonic momentum. The polar ar-PL intensity, $I(\theta)$, is depicted with the dotted orange line, where the radial distance (arrow thickness) from the source of emission indicates the intensity. (b) Excitonic band structure for  $K-K$ and $K'-K'$  excitons. Valley-exchange coupling leads to valley mixing in the excitonic dispersion resulting in  a splitting into a  parabolic and a linear band. The orthogonal circular polarised emission (orange) is mixed for both bands, while the linear and parabolic band emits only $p$ and $s$ polarised light (green), respectively. The light-cone (dashed black line) indicates the excitonic dispersion which can be probed by ar-PL.}
    \label{fig:my_label}
\end{figure}

 \paragraph*{Theoretical model:}
 A schematic outlining a typical ar-PL experiment is shown in Fig 1 (a): After unpolarised optical excitation, electron-hole pairs form at both the $K$ and $K'$ valleys in the considered MoSe$_2$ monolayer \cite{splendiani2010emerging, xiao2012coupled}. These electron-hole pairs interact via the Coulomb interaction leading to the formation of excitons.  This is  described by solving the Wannier equation \cite{kira2006many, selig2016excitonic}, providing the excitonic wavefunctions and binding energies, $E_\text{b}$.   The excitons have a parabolic dispersion $E_{\vect{Q}} =  E_\text{Gap} - E_\text{b} + \hbar^2 \vect{Q}^2/2 M$, where $\vect{Q}$ is the excitonic centre-of-mass momentum and $M$ is the centre of mass.
The valley excitons (also denoted as $K-K$ and $K'-K'$, shown in Fig. 1(b)) interact via the valley-exchange coupling,  leading to the formation of new valley-hybridised exciton states. They decay radiatively,  where the  emission angle ($\theta$) is determined by $\vect{Q}$ ensuring the conservation of  momentum.

 Taking into account the purely excitonic contributions, the exciton Hamilton operator can be written in terms of the free excitonic part and the intra/intervalley exchange  coupling \cite{qiu2015nonanalyticity, wu2017topological}
  \begin{align}
     H_X = \sum_{\vect{Q}, \xi, \xi'\neq \xi} \left(\left(\varepsilon_{\xi\vect{Q}}+\mathcal{K}^\xi_{\vect{Q}}\right) \op{X}^{\xi \dagger}_{\vect{Q}} \op{X}^{\xi}_{\vect{Q}}+  \mathcal{J}_{\vect{Q}} \op{X}^{\xi \dagger}_{\vect{Q}} \op{X}^{\xi'}_{\vect{Q}} \right), 
 \end{align}
 where $\op{X}^{\xi(\dagger)}_{\vect{Q}}$ are excitonic annihilation (creation) operators in valley $\xi$ and $\mathcal{J}_{\vect{Q}}$ and $\mathcal{K}^\xi_{\vect{Q}}$ are the  inter- and intravalley exchange matrix elements, respectively (cf. SI for more details). Equation (1) can be solved by expressing the problem in a new valley-spinor basis,
 \begin{align}
      \begin{pmatrix}
    E_{\vect{Q}} +\mathcal{K}_{\vect{Q}}  & \mathcal{J}_{\vect{Q}} \\
    \mathcal{J}^*_{\vect{Q}}&  E_{\vect{Q}} + \mathcal{K}_{\vect{Q}} 
    \end{pmatrix} \begin{pmatrix}
         U^K_{\eta,\vect{Q}} \\
          U^{K'}_{\eta,\vect{Q}}
        \end{pmatrix}
        = \mathcal{E}_{\vect{Q}, \eta}\begin{pmatrix}
             U^K_{\eta,\vect{Q}} \\
          U^{K'}_{\eta,\vect{Q}}
        \end{pmatrix}
 \end{align}
 with the energy $\mathcal{E}_{\vect{Q}, \eta} = {E}_{\vect{Q}} + \mathcal{K}^\xi_{\vect{Q}}  + \eta |\mathcal{J}_{\vect{Q}}| $, where $\eta = \pm 1$ is the new band index denoting the two split bands. For small $\vect{Q}$ within the light-cone, the upper band is approximately linear and is typically referred to as the longitudinal \cite{qiu2015nonanalyticity, schneider2020optical}  or non-analytic band \cite{wu2017topological, yu2014dirac} (right panel of Fig. 1(b)). Since $|\mathcal{J}_{\vect{Q}}| =|\mathcal{K}_{\vect{Q}}^\xi|$ (cf. SI), the lower band remains parabolic and is often called the transverse or analytic band.  In addition to the modified band structure, the exchange interaction alters the valley distribution of excitons in each band $\eta$. The terms $ U^K_{\eta,\vect{Q}} = \frac{1}{\sqrt{2}},  U^{K'}_{\eta,\vect{Q}} = \frac{\eta}{\sqrt{2}} e^{2i\theta_{\vect{Q}}}$ appearing in Eq. (2) determine the weight of each valley. Here,  $\theta_{\vect{Q}}= \arctan(\frac{Q_y}{Q_x})$ is the in-plane angle.  In order to elucidate the effect of the exchange coupling,  we focus on the case of  free-standing TMD monolayers. While a substrate will lead to screening of the Coulomb-interaction \cite{kira2006many, erkensten2021exciton}, reducing the slope of the linear band, this will partially be offset by the increased width of the light-cone due to the lower excitonic binding energy. Even for hBN encapsulation, signatures of the linear band could be resolved \cite{schneider2019shedding, schneider2020optical}.
 
 We use Heisenberg's equations of motion \cite{kira2006many} to derive the semiconductor luminescence equations \cite{kira2006many}, allowing us to obtain the  Elliot formula \cite{brem2020phonon}  for the ar-PL
 \begin{align}
     I_{\sigma}({\theta, \omega}) &=  \dfrac{2}{\hbar} \sum_\eta \text{Im}\left[\dfrac{N_{\theta, \eta}|\tilde{M}^{\sigma, \omega}_{\theta, \eta}|^2}{E_{\theta, \eta} - \hbar\omega - i(\gamma_\theta^\eta + \Gamma^\eta_\theta)}\right] 
     \end{align}
with the angle-dependent optical matrix element $\tilde{M}^{\sigma, \omega}_{\theta, \eta} = \sum_{\xi,\vect{q}, \vect{Q}} U^\xi_{\eta, \vect{Q}} M^{\xi, \sigma}_{\vect{Q}} \delta_{|\vect{Q}|, |\vect{q}| \sin\theta} \delta_{ \omega, c |\vect{q}|}$ describing the probability amplitude for a photon emitted at angle $\theta$ with optical polarisation $\sigma$. The momentum conservation   is described by the  Kronecker delta, $ \delta_{|\vect{Q}|, |\vect{q}| \sin\theta}$, which ensures that $\theta = \arcsin(\frac{|\vect{Q}_{||}|}{|\vect{q}|})$, directly relating  the exciton momentum $\vect{Q}$ and the emission angle $\theta$, particularly since the photon momentum, $|\vect{q}|$, is approximately constant within our PL energy range.    Furthermore, $\omega$ and $\sigma$ are the photon energy and the direction of the circular/linear polarisation, respectively. The second Kronecker delta relates the photon frequency to its momentum. The exciton occupation $N_{\theta, \eta}$ is approximated by the Boltzmann distribution, while the radiative and non-radiative (phonon-induced) broadening, $\gamma_\theta^\eta$ and $\Gamma_\theta^\eta$, are  calculated microscopically using a second order Born-Morkov approximation for the exciton-phonon and -photon coupling \cite{brem2020phonon} (cf. SI). Importantly, the modified excitonic dispersion and valley-mixing leads to unique behaviour in both the radiative decay rate and the phonon scattering rate.

\begin{figure*} [t!]
    \centering
    \includegraphics[width = \linewidth]{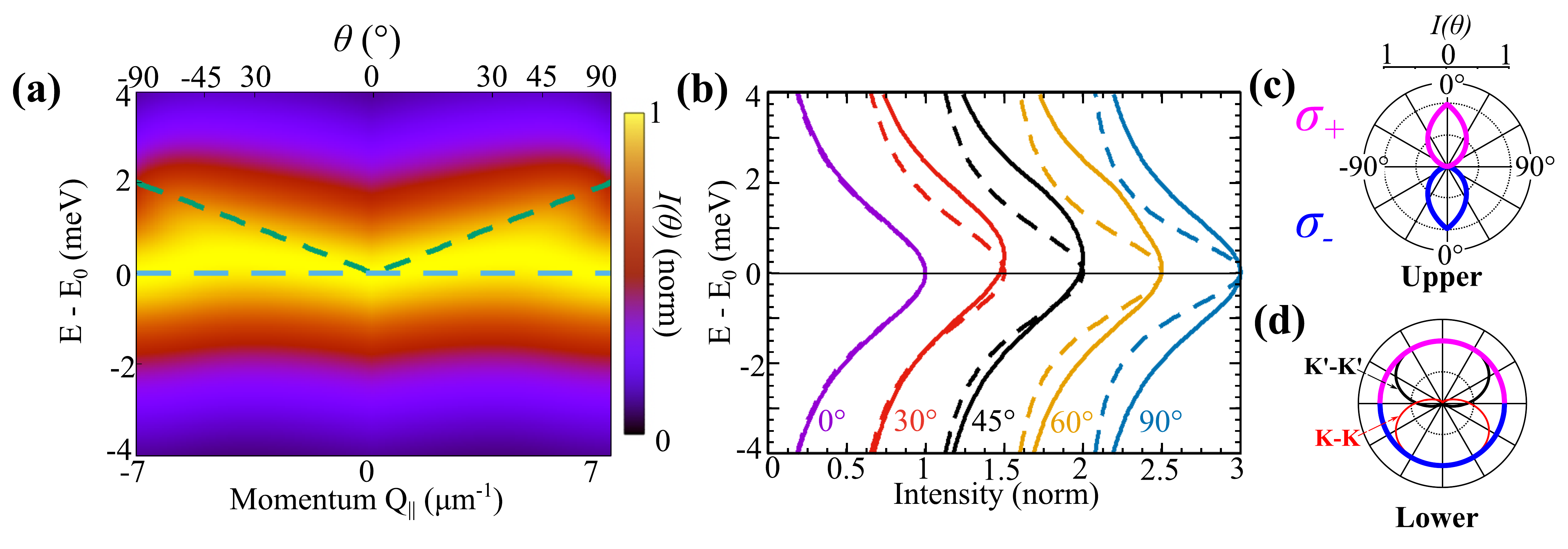}
    \caption{Angle-resolved photoluminescence. 
    (a) Normalised ar-PL spectra measured from the position of the 1s exciton peak at  $T=30$ K. The dashed lines illustrate the excitonic band structure. In our model, $0^\circ$ corresponds to emission perpendicular to the TMD plane. (b) Constant-angle cuts with (solid) and without valley-exchange coupling (dashed). Each curve is horizontally offset by 0.5 for clarity. Polar plots of ar-PL intensity $I(\theta)$ for the exchange-split upper (c) and lower (d) bands, where above and below the x-axis corresponds to $\sigma_+$ (pink) and $\sigma_-$ (blue) light emission, respectively. The polar plots  in the absence of exchange coupling are shown in (d) for the $K-K$ (thin red lines)  and the $K'-K'$ (thin black lines) exciton.}
    \label{fig:my_label}
\end{figure*}

The optical matrix element drives the angle dependence of the PL intensity. The explicit angle-dependence originates from the changing projection of the light polarisation, $\vect{e}_\sigma^\theta$, on the optical dipole of the exciton, $\vect{d}^\text{Ex}_{\vect{Q}}$, determining the matrix element  $M^{\xi, \sigma}_{\vect{Q}} = \vect{d}^\text{Ex}_{\vect{Q}}\cdot\vect{e}_\sigma^\theta $ (cf SI).  For circularly polarised light ($\sigma = \pm 1$),  the optical matrix element reads,
\begin{align}
\tilde{M}^{\sigma, \omega}_{\theta, \eta} \propto    (1+\eta) \cos \theta + \sigma (1-\eta).
\end{align}
Here, the different  angle-dependence of the two split bands becomes clear: The upper band ($\eta=+1$) has a strong $\theta$ dependence with a maximum at normal emission $\theta=0^\circ$ and zero at parallel emission $\theta=\pm90^\circ$. In contrast, the lower band ($\eta=-1$) emits with constant intensity as a function of $\theta$.
This can be understood in the following way: For a light source incident on a surface,  we  define the incidence plane as the plane perpendicular to the surface and parallel to the incident light.  Circularly polarised light can therefore be described as a superposition of $s$ and $p$ polarised light, i.e. transverse to and parallel to the incidence plane, respectively. As the angle of emission, $\theta$, increases within this plane, the $s$-component of circularly polarised light remains unchanged, while the $p$-component decreases. An exciton in valley $\xi$ has an in-plane, circular optical dipole $\vect{d}^\text{Ex}_{\vect{Q}}\propto (1, \xi i)e^{i\xi\theta_{\vect{Q}}}$, where the phase $\theta_{\vect{Q}}$ ensures that the response is isotropic.    This excitonic-dipole can also be decomposed into components parallel to or transverse to the incidence plane, which in turn couple to the $p$-component and $s$-component of the emitted photon, respectively.  Valley mixing ensures that one of these excitonic-dipole components is zero, leading to either angle-dependent emission from the upper band or angle-independent emission from the lower band. 
Importantly, the emission intensity for both bands is independent of the sign of the circular polarisation $\sigma_\pm$, resulting in equal near-orthogonal emission of both polarisations, as shown in Fig 1(b).  This situation is expected, because we are here considering the incoherent limit, where the occupation of $K$ and $K'$ valley have equilibrated after excitation or in other words an incoherent occupation of 50/50 valley hybridized states has formed.
For $s$ and $p$ linearly polarised light we obtain
$\tilde{M}^{s-pol}_{\theta, \eta}  \propto   1-\eta$ and $
    \tilde{M}^{p-pol}_{\theta, \eta}  \propto   \cos \theta (1+\eta)$, respectively.
Therefore, the upper band emits only $p$-polarised light, while the lower band emits only $s$-polarised light, i.e. the valley exchange coupling leads to a completely linear polarised emission, in stark contrast to the circularly polarised case (Fig 1(b)). 
Note that while the  optical matrix element and the polarisation of the emitted light have a strong influence on the angle dependence of the PL intensity $I_\sigma(\theta, \omega)$,  the radiative and non-radiative broadenings as well as the exciton occupation of the bands appearing in Eq. (3) will also play a role and have been explicitly considered in our microscopic calculations. 
We focus on  MoSe$_2$ monolayers primarily due to absence of low-lying dark states \cite{brem2020phonon}.  In tungsten-based TMDs these dark states have been shown to suppress the intervalley-exchange interaction \cite{selig2020suppression} In particular, we focus on MoSe$_2$ rather than MoS$_2$ because of its lower electrostatic screening. In principle however, our proposal and theory should extend to other common TMDs, such as MoS$_2$, as well as tungsten based TMDs, albeit with a weaker experimental resolution. Furthermore, theoretical studies have proposed that the presence of defects or a local modulation in the out-of-plane electric field can lead to a Rashba coupling between the bright and spin dark exciton bands \cite{dery2015polarization, yang2020exciton}. Since the Rashba effect is linearly dependent on the in-plane momentum, and scales much more slowly than the exchange interaction \cite{dery2015polarization, yang2020exciton} it has little impact on the exciton dispersion and wavefunctions within the light cone.

\paragraph*{Angle-resolved photoluminescence:}
Solving the Elliot formula [Eq. (3)], we predict the angle-resolved photoluminescence from the exchange-split upper and lower bands (Fig. 2 (a)).  The presence of the upper linear dispersion can be clearly seen. As discussed above, the optical matrix element vanishes for the upper band  for in-plane emission ($\theta=90^\circ$) leading to the linear features being diminished at large $\theta$.  To highlight this further, polar intensity maps are plotted for both bands in Figs. 2(c) and (d), respectively. Here,  the radius of the pink/blue contours corresponds to the intensity of $\sigma_-$ and $\sigma_+$ emission. The upper band  emits with maximum intensity at $\theta=0^\circ$, while at $\theta=90^\circ$ the intensity is zero. The teardrop shape of the upper band  at low angles is due to the changing exciton occupation  $N_{\theta, \eta=+1} = e^{-|\mathcal{J}\sin(\theta)|/k_B T}$ in the linear upper band, which decays rapidly at larger angles (i.e. larger in-plane momenta). The phonon-induced exciton linewidth, $\Gamma^\eta_\theta$, is highly dependent on temperature, however within the relevant momentum range it is approximately constant for both bands. 

In contrast to the strong angle-dependence for the upper band,  the emission profile for the lower parabolic band is practically constant (Fig. 2(d)), arising from an angle-independent optical matrix element, nearly constant exciton occupation, limited variation of the exciton-phonon scattering rates within the light cone, and a constant radiative decay rate, where the latter scales with the square of the optical matrix element (see supplementary) . Importantly, since the excitonic wave function is completely valley-mixed, there is no difference in the emission spectra for  $\sigma_+/\sigma_-$ polarised light.  This is distinct from the case without valley-exchange (thin red/black lines in Fig 2 (d)), where strong (and opposite) optical polarisation dependence is observed. In particular, at $\theta = 0^\circ$, the $K-K$ ($K'-K'$) exciton emits only $\sigma_-$ ($\sigma_+$) photons. However, since these two exciton species would be equally occupied, due to their degeneracy, the resulting emission would also be unpolarised.  In contrast, at $90^\circ$, the projection of both polarisations to the in-plane excitonic dipoles are equivalent leading to equal PL intensities.

The angle-dependent PL from the upper band can be seen more clearly in Fig 2(b), where for $\theta = 30^\circ$, $45^\circ$ and $60^\circ$ show an asymmetry in the intensity profile. For comparison, the same normalised ar-PL intensity cuts are shown in the absence of valley-exchange coupling with dashed lines. They exhibit clearly symmetric lines for all angles.   This demonstrates that the ridge-like feature should be a clear experimental signature of the exchange interaction. The curves at $\theta = 0^\circ$ are very similar because the valley-exchange vanishes at $\vect{Q}_{||} = 0$, however they differ slightly because the exciton phonon-scattering rate depends on the entire excitonic band structure. At $90^\circ$, even though the upper band contribution is 0, leading to a symmetric intensity profile, the constant optical matrix element originating from the exchange interaction leads to a larger overall radiative broadening. Note that while we assume an unpolarised optical excitation for simplicity, the same equal occupation in both valleys can be achieved via linear polarisation. Furthermore, the Coulomb-exchange mechanism has been proposed as a source of valley decoherence in TMD monolayers \cite{hao2016direct}, such that even under circularly polarised excitation, this valley-exchange distribution can be created. Similarly, relaxation processes within the light cone ensure that the exciton distribution is independent of the excitation-incidence angle, and can be described by a Boltzmann distribution \cite{brem2018exciton}.

 \paragraph*{Tuning of angle-resolved PL in a magnetic field}
 
So far, we have demonstrated that the valley-exchange interaction leads to an even mixing of valley-states in both exchange-split exciton bands. However,  the valley distribution of each band can be tuned by means of an external magnetic field, offering an additional experimental probe of the exchange interaction.
The alignment of particle spin to an applied magnetic field leads to a Zeeman energy shift. In a TMD, the coupled valley-spin degree of freedom leads to an effective valley splitting of  excitons  \cite{li2014valley, srivastava2015valley, van2018strong, koperski2018orbital}, where the $K$ and $K'$ valley undergo equal and opposite energy shifts. We add the valley Zeeman terms in Eq. (2) and find that the linear and parabolic bands become modified resulting in 
\begin{align} \label{mag}
  \mathcal{E}_{\vect{Q}, \eta} (B)= {E}_{\vect{Q}} + \mathcal{K}_{\xi\vect{Q}}  + \eta\sqrt{(g_L\mu_B B)^2 + |\mathcal{J}_{\vect{Q}}|^2}  
\end{align} 
where $g_L$ and $\mu_B$ are the g-factor and Bohr magneton, respectively, with $g_L\mu_B = 0.24 $meV/T \cite{koperski2018orbital}. 
For large magnetic fields, the third term in Eq. (\ref{mag}) is approximately constant in momenta, leading to both bands becoming  quasi-linear, as shown by the dashed lines in Fig 3(a).  Furthermore, the magnetic field modifies the valley-distribution in each band, which now becomes strongly $\vect{Q}$ dependent with (cf. SI), $ U^{K}_{\eta,\vect{Q}} =  \frac{1}{\sqrt{1+|\alpha^{\eta}_{\vect{Q}}|^2}}$ and $ U^{K'}_{\eta,\vect{Q}} = \frac{\alpha^{\eta}_{\vect{Q}} e^{2i\theta_{\vect{Q}}}}{\sqrt{1+|\alpha^{\eta}_{\vect{Q}}|^2}}$.
Here, we have introduced the term $\alpha^\eta_{\vect{Q}} = (g_L \mu_B B  +\eta \sqrt{(g_L \mu_B B)^2 +|\mathcal{K}_{\vect{Q}}|^2 })/(|\mathcal{K}_{\vect{Q}}|)$, which determines the valley-distribution at momentum $\vect{Q}$ in band $\eta$.

\begin{figure}[t!]
    \centering
    \includegraphics[width = 0.99\linewidth]{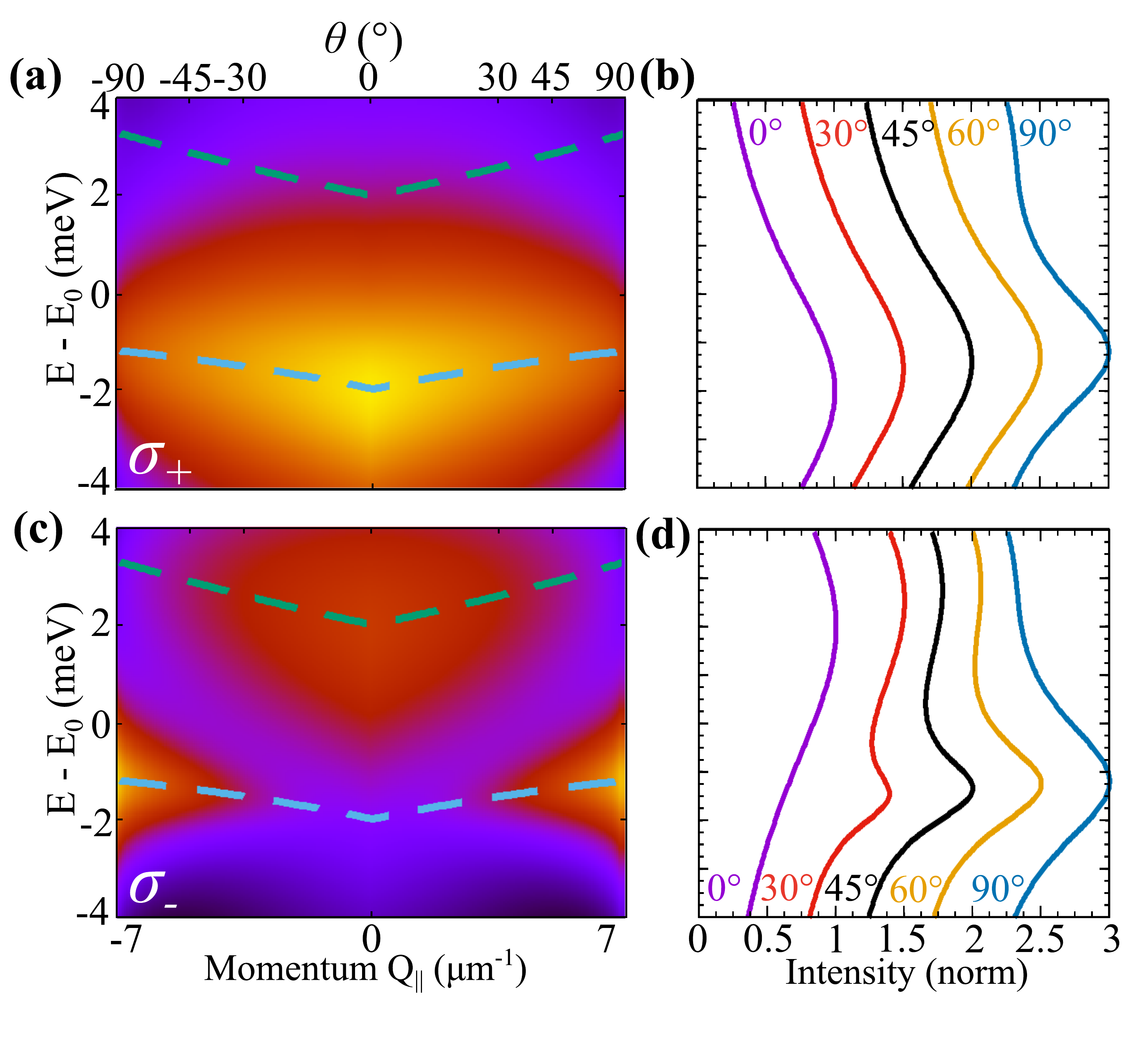}
    \caption{Angle-resolved PL in magnetic field.   Normalised ar-PL spectra measured from the position of the 1s-exciton peak for (a) $\sigma_+$  and (c) $\sigma_-$ at $B=8$ T and $T = 77$ K.  The dashed lines denote the excitonic band structure. Same colour scale as in Fig 2.  (b,d) Normalised constant angle cuts, where each curve is horizontally offset by 0.5 for clarity. }
    \label{Fig1}
\end{figure}

In Figs.3 (a) and (c),  we show the ar-PL intensity at $T=77$ K and in the presence of a  $B=8 $T perpendicular magnetic field for $\sigma_+$ and $\sigma_-$ emission polarisation, respectively. Unlike in Fig 2 showing the ar-PL in absence of magnetic fields, $\sigma_+$ and $\sigma_-$ emission have clearly distinct PL maps. The magnetic field has two effects: (i) It opens an energy gap at $Q_{||}=0$ as the K and K' valleys in a TMD carry different spin-polarisations, cf.  the dashed lines in Fig 3 (a) and (c). The upper and lower bands are now both quasi-linear. (ii) The magnetic field changes the valley mixing within each band. At small $\vect{Q}_{||}$,  $\mathcal{K}_{\vect{Q}}\ll g_L \mu_B B$, so the magnetic field leads to  completely valley polarised energy bands. This is clear as $\alpha_{\vect{Q}}^\eta \sim (1+\eta) g_L\mu_B B/\mathcal{K}_{\vect{Q}}$ becomes large with strong $\eta$ dependence. This is evident in Figs. 3(a) and (c), where the lower band emits only $\sigma_+$ polarised light, while the upper band only emits $\sigma_-$ light. As $\vect{Q}_{||}$ ($\theta$) increases, the bands become less valley-polarised and both $\sigma_+$ and $\sigma_-$ photons start to be emitted from both bands. Similarly to the case without magnetic field, the rotation of the photon emission vector modifies the optical matrix element, such that $\sigma_-$ and $\sigma_+$ polarised light are emitted with equal intensity at $90^\circ$ . In Fig. 3(a), this leads - combined with the thermal exciton occupation -  to a negligible emission from the upper band. In contrast, Fig. 3 (c) demonstrates that for $\sigma_-$ emission, a transition occurs at a certain angle. The vanishing optical matrix element of the lower band near $Q_{||}= 0$ leads to a dominant optical response from the upper band. As the momentum/emission angle increases the optical matrix elements become similar, such that due to the small thermal exciton occupation of the upper band, the lower band becomes dominant.
The difference in the PL emission profiles can be more easily seen in the normalised intensity cuts in Fig 3(d): At $0^\circ$, the emission from the upper band is dominant, at $30^\circ$ both bands emit similarly, while at $60^\circ$ the  lower band dominates the PL. 

\begin{figure}[t!]
    \centering
    \includegraphics[width = 0.99\linewidth]{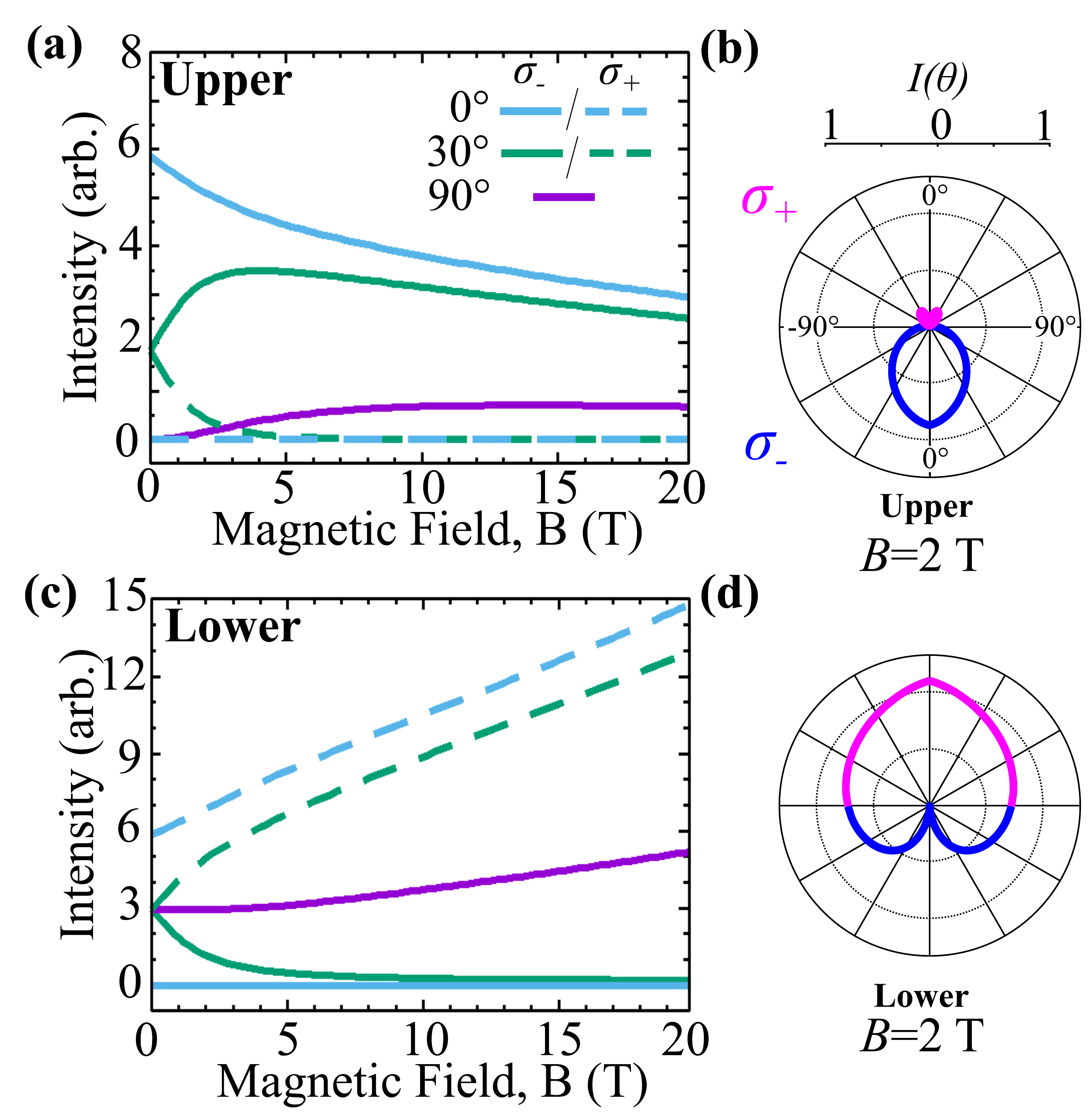}
    \caption{Magnetic field dependence of the angle-resolved PL for the (a) upper and the (c) lower band. The solid (dashed) lines correspond to $\sigma_-$ ($\sigma_+$) polarisation.  Polar plots of ar-PL intensity $I(\varphi, \theta)$ for the exchange split (b) upper  and (d) lower band at $T=100$ K and $B=2$ T.  Here, the line above and below the x-axis corresponds to $\sigma_+$ (pink) and $\sigma_-$ (blue) emission, respectively. }
    \label{Fig1}
\end{figure}

To further explore the effect of the magnetic field, we show in Fig. 4 the intensity of the upper and lower band at a fixed temperature $T=100$ K and at given angles as a function of the magnetic field for $\sigma_-$ (solid) and $\sigma_+$ (dashed) emission.   We find a decrease in the emission intensity of the upper band with an increasing magnetic field, as the Zeeman-induced band gap leads to a lower exciton occupation, cf.  the solid $\theta = 0^\circ$ line in Fig. 4(a). The dashed line for $\theta = 0^\circ$ denoting the $\sigma_+$ polarisation is zero due to the valley selection rule. At $90^\circ$, both $\sigma_-$ and $\sigma_+$ are equivalent. As discussed above,  the intensity from the upper band vanishes at $90^\circ$ at zero magnetic field owing to the optical matrix element. However, the latter depends on the magnetic field and the valley occupation becomes more and more polarised with an increasing  field allowing for a larger emission in the plane, cf. Eq. (8). This competes with the intensity loss from the lower thermal occupation from the Zeeman splitting giving rise to a maximum occurring around 9 T for $\theta = 90^\circ$.  For $\theta =30^\circ$ the bump occurs earlier (3 T) as for a given magnetic field, the valley polarisation is larger at smaller momenta $\vect{Q}_{||}$. For $\sigma_+$ emission the intensity at $\theta =30^\circ$ decays rapidly, as the upper band becomes more polarised. 

For the lower exchange-split band,  we find  a qualitatively different behaviour, cf. Fig 4 (c). Since the Zeeman effect shifts the lower band in the opposite direction, the thermal exciton occupation of this band becomes larger as the magnetic field increases. This leads to an overall increased intensity from the lower band as reflected in the larger y-axis scale. At $\theta=0$, the lower band is completely valley polarised, leading to a maximum in the $\sigma_+$ and a minimum in the $\sigma_-$ emission, opposite to situation in the upper band. For all non-zero angles the intensity is constant at zero magnetic field. As the field increases, the lower band becomes more valley-polarised leading to an enhanced $\sigma_+$  and a reduced $\sigma_-$ emission at $\theta=30^\circ$. At $90^\circ$, both $\sigma_-$ and $\sigma_+$ are equivalent as discussed above. Interestingly, magnetic-field-induced reduction in the exciton-phonon scattering rates also leads to an increase in ar-PL intensity. This is reflected in the behaviour of the dashed $\theta = 0^\circ$ line, which  increases linearly due to the competition between the exponentially increasing exciton occupation and a reduction in the phonon-induced dephasing.

 We show  characteristic polar intensity plots for a fixed magnetic field $B=2$ T and temperature $T=100 K$ to elucidate the impact of the magnetic field on $\sigma_+$/$\sigma_-$ emission, cf. Figs. 4(b) and (d) illustrating the upper and the lower band, respectively.   For near-orthogonal emission, $\sigma_-$ ($\sigma_+$) is favoured in the upper (lower) band, as discussed above. Furthermore, for emission within the plane, $\sigma_-$ and $\sigma_+$ polarisations are equivalent as demonstrated in both the upper and lower band. The Boltzmann distribution at 100 K strongly diminishes the overall emission intensity from the upper band due to the field-enhanced energy offset.  
 Overall, an external magnetic field allows the valley polarisation of the bands to be continuously tuned offering the possibility to control the ar-PL spectra. 
  
  \begin{figure}[t!]
    \centering
    \includegraphics[width = 0.95\linewidth]{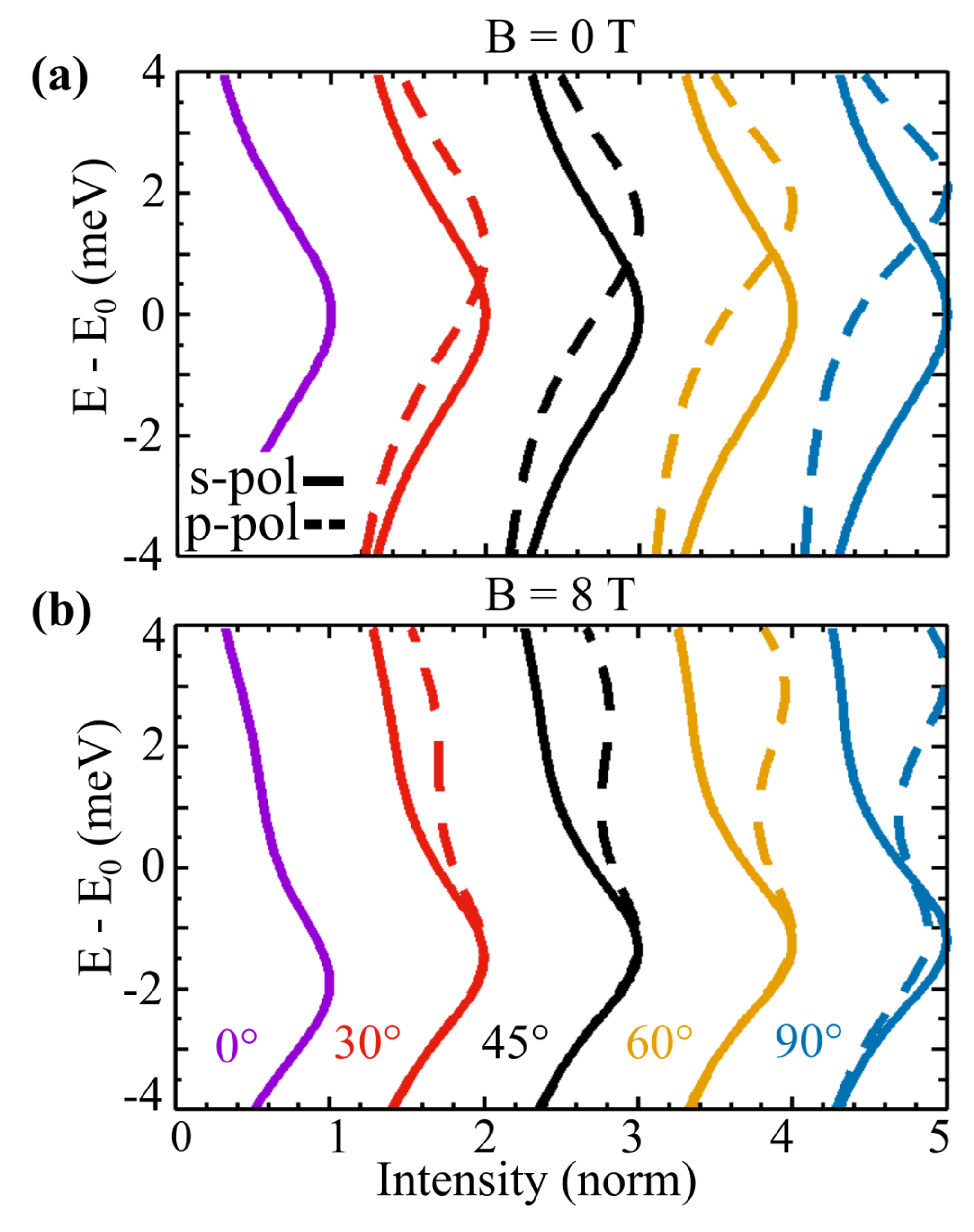}
    \caption{Angle-Resolved PL  for (a) $B=0$ T and (b) $B=8$ T at $T = 77$ K for $s$ (solid) and $p$ (dashed) polarisation at constant angle cuts. Curves for each angle are offset by 1 for clarity.}
    \label{Fig1}
\end{figure}

\paragraph*{Linear angle-resolved PL in a magnetic field:}
So far, we have demonstrated that valley-exchange driven effects can be probed by an angle-resolved PL experiment involving circularly polarised light. However, similar experiments \cite{wang2017plane, schneider2019shedding} have been performed using linearly polarised light.
The two exciton branches formed as a result of exchange-splitting are often referred to as the longitudinal and transverse branches \cite{qiu2015nonanalyticity} reflecting their distinct coupling to linearly polarised light. In particular, it has been predicted that the upper, longitudinal branch couples only to $p$-polarised light, whose polarisation is parallel to the the incidence plane, while the  lower transverse branch couples only to $s$-polarised light, with polarisation perpendicular to the incidence plane \cite{qiu2015nonanalyticity}.  As in the circularly-polarised case, the ar-PL intensity strongly depends on the optical matrix element.

In Fig. 5, we show the normalised ar-PL intensity for a range of angles for $B=0$ T and $B=8$ T. The solid (dashed) lines correspond to $s$-polarised ($p$-polarised) emission, such that in (a) the lower (upper) band is probed independently. In general, the intensity of the upper band is small, due to low thermal exciton occupation.  However, since $p$-polarisation completely diminishes the emission from the lower branch, the linear dispersion of the upper branch is more easily probed than in the circularly polarised case. 
For $B=8$ T, the magnetic response is less striking than in the circularly polarised case. While, $s$-polarisation still probes the lower band, the magnetic-field-induced valley polarisation of the bands relaxes the linearly-polarised selection rule. Hence, for $\theta \sim 0^\circ$ the PL intensity profiles are the same for both polarisations. As before, valley-mixing grows as the exciton momentum/emission angle increases recovering the $s$/$p$ selection rules. While this leaves the $s$-polarised emission more or less independent of emission angle as it is dominated by the lower branch, the competition between valley mixing and thermal occupation results in similar emission from both branches for $p$-polarisation at larger angles, as seen in the double bumps in Fig 5(b). 
Overall, measurements of both linear and circularly polarised angle-resolved PL intensity should allow the observed excitonic dispersion to be unambiguously attributed to the valley-exchange splitting.

In conclusion, we have demonstrated that angle-resolved photoluminescence can be used to determine the impact of valley-exchange coupling in  monolayer TMDs.   We calculate the change of exciton dispersion due to valley-exchange  and determine microscopically the changes in excitonic linewidths including the radiative recombination and exciton-phonon scattering. We demonstrate that the non-analytic linear dispersion is clearly distinguishable in the angle-resolved PL spectra. Furthermore, the effects of a magnetic field are discussed which offers an alternative probe of the valley-exchange coupling strength as well as a clear way to distinguish the two exchange-split bands. 
We also show that linearly polarised emission  allows the distinct exciton bands to be probed independently. A combined measurement of both linearly and circularly polarised light should allow both the excitonic dispersion and the valley projections on both bands to be determined in the experiment. Overall, our work provides a microscopic understanding of the impact of the altered exciton dispersion due to valley-exchange coupling on the optical properties of atomically thin semiconductors.

\begin{acknowledgements}
This project has received funding from the Excellence Initiative Nano at Chalmers and the Knut and Alice Wallenberg Foundation via the Grant KAW 2019.0140. W.W. acknowledges support by the Knut and Alice Wallenberg foundation through a Wallenberg Academy Fellowship (No. 2019.0231) and by the Carl Tryggers foundation (CTS 19:406).  Furthermore, we are thankful to the Deutsche Forschungsgemeinschaft (DFG) via CRC 1083 and the European Unions Horizon 2020 research and innovation programme under grant agreement no. 881603 (Graphene Flagship). We acknowledge the Vinnova competence centre ``2D-TECH". C.S. gratefully acknowledges support of the European Research Council (ERC) within the project UnLimit2D (project no. 679288), and the DFG within project SCHN1376 11.1. 
\end{acknowledgements}

\bibliographystyle{rsc}
\renewcommand*{\citenumfont}[1]{#1}
\renewcommand*{\bibnumfmt}[1]{[#1]}
\providecommand{\noopsort}[1]{}\providecommand{\singleletter}[1]{#1}%
\providecommand*{\mcitethebibliography}{\thebibliography}
\csname @ifundefined\endcsname{endmcitethebibliography}
{\let\endmcitethebibliography\endthebibliography}{}

\end{document}